\documentclass[letterpaper]{article}
\usepackage[T1]{fontenc}

\usepackage{geometry}
\usepackage{setspace}

\usepackage[journal=jacsat,articletitle=true]{achemso}
\usepackage{graphicx}
\usepackage{float}
\newfloat{scheme}{htbp}{los}
\floatname{scheme}{Scheme}
\floatname{chart}{Chart}
\newfloat{graph}{htbp}{loh}
\usepackage{siunitx}
\usepackage{xcolor}

\newcommand{\eg}{{\it e.g.}, }
\newcommand{\ie}{{\it i.e.}, }

\usepackage{chemformula} % Formulas using \ch{}
\usepackage[version = 4]{mhchem} % Formulas using \ce{}

\newcommand{\exciting}{{\usefont{T1}{lmtt}{b}{n}exciting}}

\usepackage{xr-hyper}
\usepackage{amsmath}
\usepackage[hidelinks]{hyperref}

\usepackage{authblk}
\author[1]{Lu Qiao*}

\affil[1]{Department of Physics and CSMB, Humboldt-Universit\"at zu Berlin, Berlin, Germany}

\author[2]{Ronaldo Rodrigues Pela}
\affil[2]{Distributed Algorithms and Supercomputing Department, Berlin, Germany}

\affil[3]{European Theoretical Spectroscopy Facility (ETSF)}
\author[1,3]{Claudia Draxl}

\title{Disentangling Electronic and Lattice Contributions to Transient Absorption in Metal Halide Perovskites: A First-Principles Study of \ce{CH3NH3PbBr3}}

\date{*Email: qiaolu@physik.hu-berlin.de}

\begin{document}

\maketitle

\begin{abstract}

Soft lattices combined with strong electron-phonon coupling in metal halide perovskites result in a complex interplay between electronic and lattice degrees of freedom. This interplay complicates the interpretation of time-resolved excitation spectra like pump-probe spectra. Here, we develop a first-principles approach that combines a nonequilibrium extension of the Bethe-Salpeter equation with \textit{ab initio} molecular dynamics to resolve the origin of transient absorption. This approach can quantitatively disentangle electronic and thermal lattice contributions across femtosecond-to-picosecond timescales. Exemplified with \ce{CH3NH3PbBr3}, we find that on the femtosecond scale, both X-ray and optical transient absorption spectra are dominated by electronic contributions: Photoinduced Coulomb screening weakens the effective electron-hole interaction and blueshifts the excitonic resonances, whereas Pauli blocking is negligible. On the picosecond scale, thermal lattice contributions become essential, with distinct mechanisms dominating different spectral regions: Lattice vibrations lead to spectral redistribution in the X-ray transient absorption spectrum, whereas lattice expansion blueshifts the optical transient absorption spectrum.
 
\end{abstract}

\section{Keywords}
Transient absorption spectrum, electronic and lattice contributions, nonequilibrium Bethe-Salpeter equation, \textit{ab initio} molecular dynamics, metal halide perovskites

%%%%%%%%%%%%%%%%%%%%%%%%%%%%%%%%%%%%%%%%%%%%%%%%%%%%%%%%%%%%%%%%%%%%%%%%%%%%%%%%%%
%%%%%%%%%%%%%%%%%%%%%%%%%%%%%%%% Introduction %%%%%%%%%%%%%%%%%%%%%%%%%%%%%%%%%%%%
%%%%%%%%%%%%%%%%%%%%%%%%%%%%%%%%%%%%%%%%%%%%%%%%%%%%%%%%%%%%%%%%%%%%%%%%%%%%%%%%%%
\section{Introduction}

Lead halide perovskites have emerged as promising materials for optoelectronic applications due to their excellent light absorption and emission properties \cite{kojima2009,shoaib2017,liu2021}. A microscopic understanding of these properties requires insight into their ultrafast excitation dynamics, which can be probed by pump-probe transient absorption spectroscopy \cite{maiuri2019,buzzi2018}. However, the transient response arises from intertwined contributions of photoexcited carriers and thermal lattice effects \cite{droseros2024,manser2014,smejkal2020,guo2018,wang2023,wang2024}, whose individual roles remain poorly understood. The spectral features in the optical transient absorption have been commonly attributed to many-body electronic effects, such as phase-space filling due to Pauli blocking or screening by photoexcited carriers \cite{droseros2024,manser2014,smejkal2020}. However, thermal lattice effects, including lattice expansion and vibrations, can also reshape the transient spectral lineshape \cite{guo2018,wang2023,wang2024}. Although a recent experiment separated the electronic and lattice contributions to the optical transient absorption of \ce{CH3NH3PbBr3} \cite{wang2024}, the microscopic origins underlying these contributions remain unresolved. In particular, the overlapping spectral signatures of Pauli blocking, photoinduced Coulomb screening, lattice expansion and vibrations make it difficult to determine their individual roles from experiment alone. A first-principles investigation is therefore required to resolve their respective origins.

Several theoretical methods have been developed to simulate transient absorption spectra. Early density-matrix treatments of transient absorption often relied on model Hamiltonians and phenomenological relaxation and dephasing parameters, and were therefore not fully from first principles~\cite{lindberg1988,pollard1990,yan1997,wolfseder1997}. Real-time time-dependent density-functional theory (RT-TDDFT) has been widely used to simulate transient absorption (TA) spectra~\cite{de2013,pemmaraju2020,moitra2023}. However, electron-hole interactions within the RT-TDDFT formalism are approximated through the exchange-correlation kernel, making the description of many-body effects strongly dependent on the choice of kernel. Many-body wavefunction-based methods, including time-dependent configuration-interaction~\cite{pabst2012} and coupled-cluster approaches~\cite{skeidsvoll2020}, provide an explicit treatment of electronic correlation. However, their high computational cost limits their applications to atoms and small molecules. The Bethe-Salpeter equation (BSE) is the state-of-the-art approach to describing electron-hole interactions in solids~\cite{onida2002,sondersted2024,schebek2025,alvarez2023}. Its conventional formulation, however, relies on ground-state carrier distributions and therefore cannot describe the modified electronic populations and screening induced by photoexcitation. To address this limitation, a nonequilibrium BSE formalism was derived for simulating the optical response of photoexcited systems \cite{perfetto2015}. Based on this formalism, we recently developed a first-principles approach that incorporates photoexcited carrier distributions into the BSE within the adiabatic approximation. We have implemented this approach in the all-electron code \exciting~\cite{gulans2014,raya2026} and applied it to several semiconductors, successfully reproducing their experimental transient spectra and separating electronic contributions from those induced by lattice expansion \cite{rossi2025,qiao2025}.

In this work, we go beyond the existing formalism by combining it with \emph{ab initio} molecular dynamics (AIMD) to incorporate thermal effects from lattice vibrations. We apply our approach to simulate the TA spectra of \ce{CH3NH3PbBr3}, a versatile perovskite used in applications ranging from solar cells to X-ray detectors~\cite{kojima2009,edri2013,wei2016,liu2024}. We distinguish between electronic contributions (photoinduced Coulomb screening and Pauli blocking) and lattice contributions (lattice expansion and vibrations), and determine how each shapes the transient response. We find that the femtosecond transient absorption is dominated by photoinduced Coulomb screening in both the X-ray and optical regimes. On the picosecond timescale, the dominant contributions depend on the energy window: X-ray transient absorption is governed by lattice vibrations together with photoinduced Coulomb screening, whereas optical transient absorption mainly arises from lattice expansion and Pauli blocking.

%%%%%%%%%%%%%%%%%%%%%%%%%%%%%%%%%%%%%%%%%%%%%%%%%%%%%%%%%%%%%%%%%%%%%%%%%%%%%%%%%%
%%%%%%%%%%%%%%%%%%%%%%%%%%%%%%%%%%%% Method %%%%%%%%%%%%%%%%%%%%%%%%%%%%%%%%%%%%%%
%%%%%%%%%%%%%%%%%%%%%%%%%%%%%%%%%%%%%%%%%%%%%%%%%%%%%%%%%%%%%%%%%%%%%%%%%%%%%%%%%%
\section{Method}

%%%%%%%%%%%%%%%%%%%%%%%%%%%%%%%%%%%%%%%%%%%%%%%%%%%%%%%%%%%%%
%%%%%%%%%%%%%%%%%%  Electronic contribution  %%%%%%%%%%%%%%%%
%%%%%%%%%%%%%%%%%%%%%%%%%%%%%%%%%%%%%%%%%%%%%%%%%%%%%%%%%%%%%

\begin{figure}
    \centering
    \includegraphics[width=1\linewidth]{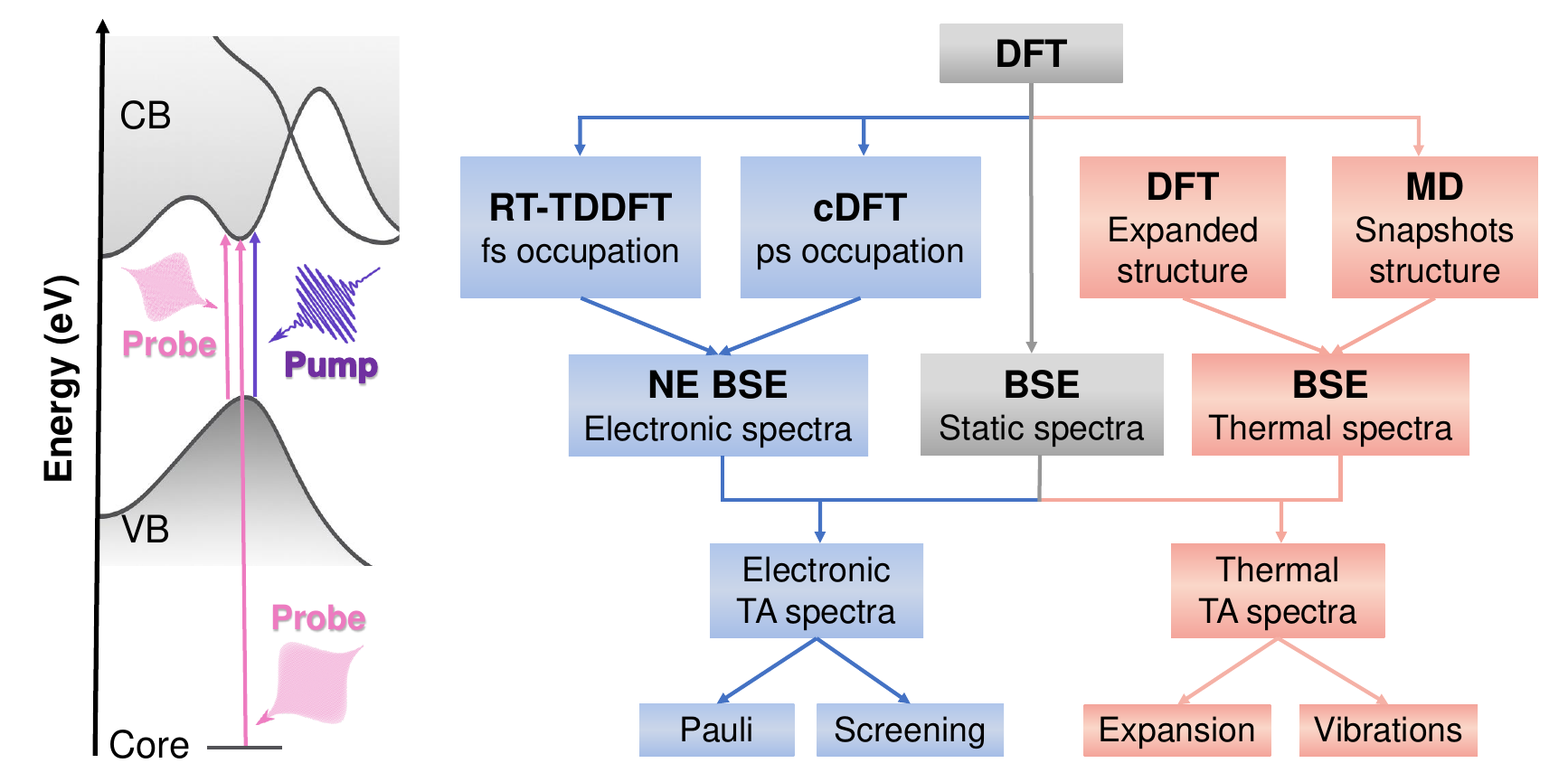}
    \caption{Left: Schematic illustration of the pump-probe process. A pump pulse photoexcites the semiconductor, initiating transitions from valence to conduction states (purple arrow). A time-delayed probe pulse subsequently measures the transient response by promoting electrons from either core or valence states to conduction states (pink arrows). Right: Computational workflow for disentangling electronic and lattice contributions to the transient spectra. \textbf{NE BSE} refers to the nonequilibrium extension of the BSE. \textbf{Pauli} and \textbf{Screening} refer to photoinduced Pauli blocking and Coulomb screening effects, respectively. \textbf{Expansion} and \textbf{Vibrations} refer to lattice expansion and vibrations effects, respectively.}
    \label{fig:workflow}
\end{figure}

To disentangle the electronic and lattice contributions to the TA spectrum, RT-TDDFT and constrained density functional theory (cDFT) are used to describe the electronic contribution induced by pump excitation, whereas \textit{ab initio} molecular dynamics (AIMD) is employed to capture the lattice contribution. The subsequent probe process is simulated within \exciting's nonequilibrium BSE framework, which yields the optical response of the photoexcited system. Figure~\ref{fig:workflow} illustrates the pump-probe process and the computational workflow.

The electronic contribution accounts for the effect of photoexcited carriers on the spectrum. In \ce{CH3NH3PbBr3}, photoexcitation initially generates a hot-carrier distribution, which subsequently relaxes toward the band-edge region on timescales of $\sim$\SI{300}{femtoseconds}~\cite{richter2017}. To describe the hot-carrier distribution at early stage, we employ RT-TDDFT to propagate the carrier populations by projecting the time-evolved Kohn-Sham orbitals onto the ground-state orbitals \cite{RodriguesPela2021,pela2024}. On the picosecond timescale, the photoexcited carriers are assumed to reach a quasi-equilibrium distribution, approximated by a Fermi-Dirac distribution described within cDFT. The photoexcited carrier distributions are then incorporated into the BSE to compute the nonequilibrium spectrum, where they affect the spectrum through two many-body mechanisms: (i) \emph{Pauli blocking}, arising from the suppression of optical transitions by state filling, and (ii) \emph{photoinduced Coulomb screening}, arising from the carrier-induced modification of the electron-hole interaction. These two effects are disentangled by selectively introducing the photoexcited carrier distributions either into the transition coefficients (Eq.~S12) to isolate \emph{Pauli blocking}, or into the statically screened Coulomb interaction (Eq.~S9-S10) to isolate \emph{photoinduced Coulomb screening}. The difference between the electronic and static absorption spectra reflects the many-body electronic effects on the transient response.  

\newpage

The lattice contribution to the spectrum of \ce{CH3NH3PbBr3} is evaluated at delays of several hundred picoseconds, where experiments have shown that the lattice response becomes substantial~\cite{wang2024,guzelturk2021}. Photoexcitation modifies the lattice through both structural expansion and atomic vibrations. The former is modeled by expanding the lattice parameters~\cite{qiao2025}, whereas the latter is described using \textit{ab initio} molecular dynamics (AIMD). To account for vibrational effects, we average the absorption spectra calculated from representative AIMD snapshots. By comparing the thermal spectrum with the static spectrum, we isolate the lattice contribution to the transient spectrum. The formalism and computational details are summarized in Sections~S1 and S2 of the Supporting Information.

%%%%%%%%%%%%%%%%%%%%%%%%%%%%%%%%%%%%%%%%%%%%%%%%%%%%%%%%%%%%%%%%%%%%%%%%%%%%%%%%%%
%%%%%%%%%%%%%%%%%%%%%%%%%%%   Result and discussion  %%%%%%%%%%%%%%%%%%%%%%%%%%%%%
%%%%%%%%%%%%%%%%%%%%%%%%%%%%%%%%%%%%%%%%%%%%%%%%%%%%%%%%%%%%%%%%%%%%%%%%%%%%%%%%%%
\section{Results and discussion}
%%%%%%%%%%%%%%%%%%%%%%%%%%%%%%%%%%%%%%%%%%%%%%%%%%%%%%%%%%%%%
%%%%%%%%%%%%%  XTA spectra at the Br K-edge on the fs  %%%%%%%%%%
%%%%%%%%%%%%%%%%%%%%%%%%%%%%%%%%%%%%%%%%%%%%%%%%%%%%%%%%%%%%%

\subsection{XTA spectra at the Br K-edge on the femtosecond timescale}

\begin{figure}[H]
    \centering
    \includegraphics[width=0.5\linewidth]{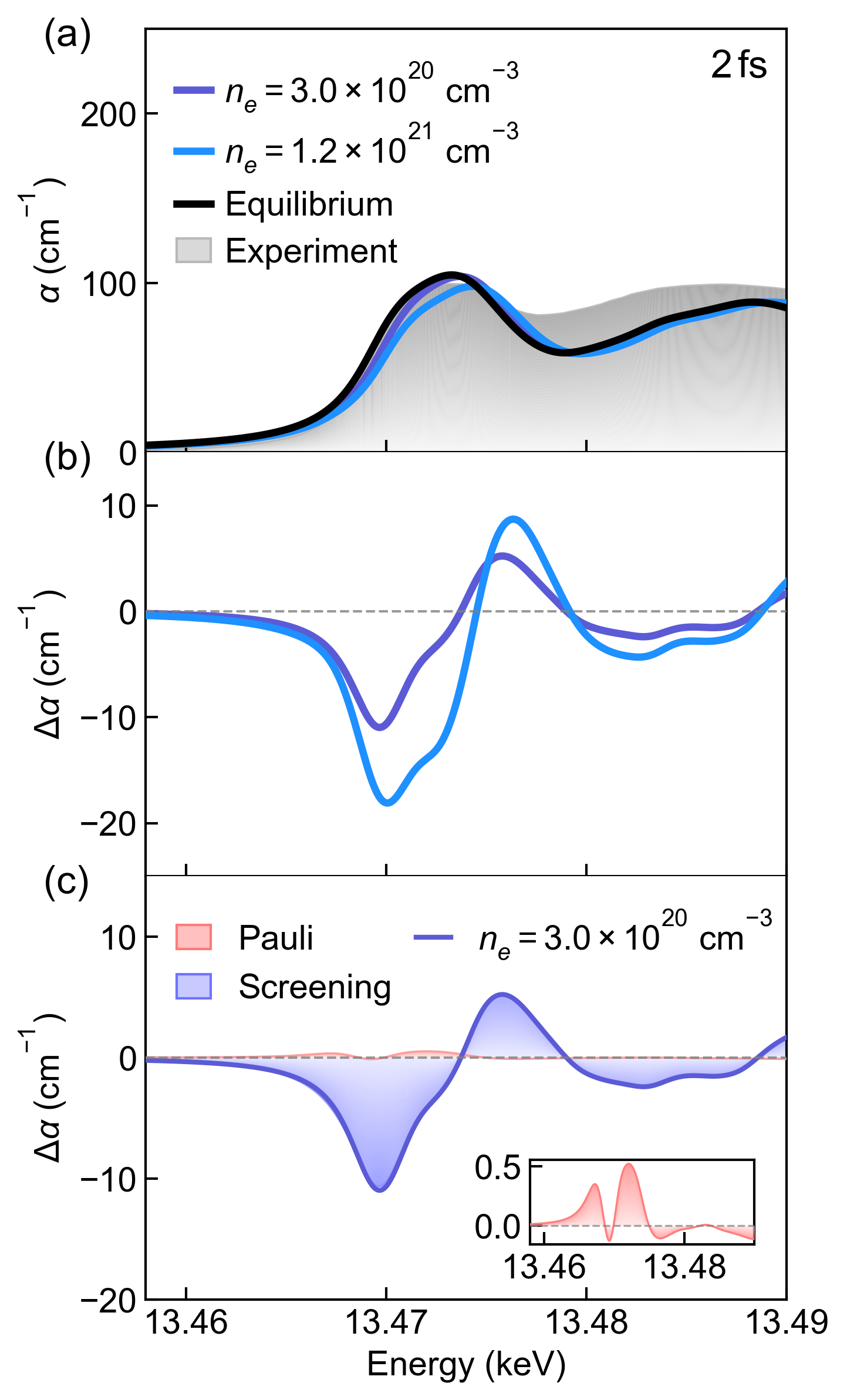}
    \caption{(a) Comparison of equilibrium (black curve) and nonequilibrium (blue curves) X-ray absorption spectra (XAS) at the Br K-edge of \ce{CH3NH3PbBr3} at a time delay of \SI{2}{\femto\second}, showing the electronic contributions at different excitation densities $n_e$. The experimental spectrum (gray area) is taken from Ref.~\cite{liu2020}. (b) Corresponding X-ray transient absorption (XTA) spectra at the same time delay. (c) Decomposition of the XTA spectrum at $n_{e}=3.0\times10^{20}$ cm$^{-3}$ into contributions from photoinduced Coulomb screening (blue area) and Pauli blocking (red area). Inset: Zoom into the Pauli blocking contribution.}
    \label{fig:XTA_fs}
\end{figure}

We first look at the femtosecond timescale, including only electronic contributions. Figure~\ref{fig:XTA_fs}a shows the calculated equilibrium and nonequilibrium X-ray absorption spectra (XAS) at a time delay of \SI{2}{\femto\second}. The equilibrium spectrum (black curve) shows good agreement with the experimental counterpart (gray area) \cite{liu2020}. Upon photoexcitation, the nonequilibrium spectra (blue curves) show a blueshift accompanied by a redistribution of spectral weight. The blueshift of the peak at $\sim$\SI{13.474}{\kilo\electronvolt} increases from \SI{0.42}{\electronvolt} at an excitation density of $n_{e}=3.0\times10^{20}$ cm$^{-3}$ to \SI{1.05}{\electronvolt} at $n_{e}=1.2\times10^{21}$ cm$^{-3}$. This blueshift gives rise to the X-ray transient absorption (XTA) shown in Fig.~\ref{fig:XTA_fs}b, which consists of a negative feature at $\sim$\SI{13.470}{\kilo\electronvolt} and a positive feature at $\sim$\SI{13.478}{\kilo\electronvolt}. The absence of an isosbestic point at $\sim$\SI{13.475}{\kilo\electronvolt} indicates a non-linear increase in XTA amplitude with excitation density, \eg the XTA amplitude at $\sim$\SI{13.470}{\kilo\electronvolt} increases approximately as $\sqrt{n_{\mathrm e}}$. Such a non-linear scaling of the transient absorption amplitude with the excitation density has also been reported in photoexcited carbon nanotubes~\cite{hagen2004}.

To understand the origin of the XTA features, we decompose the transient response into contributions from photoinduced Coulomb screening (blue area) and Pauli blocking (red area), shown in Fig.~\ref{fig:XTA_fs}c. The XTA spectrum is dominated by photoinduced Coulomb screening, whereas Pauli blocking remains more than one order of magnitude smaller over the entire energy range. This decomposition clarifies the microscopic origin of the spectral blueshift in Fig.~\ref{fig:XTA_fs}a: Upon pumping, the photoexcited carriers enhance the dielectric screening, thereby reducing the effective Coulomb attraction between the electron and the core hole. The asymmetric XTA lineshape indicates an additional redistribution of spectral weight. The weaker high-energy feature at $\sim$\SI{13.478}{\kilo\electronvolt}, relative to the low-energy feature at $\sim$\SI{13.470}{\kilo\electronvolt}, shows that the blueshift is not uniform, but that the low-energy exciton is more sensitive to photoinduced screening than high-energy excitations. The blueshift and redistribution of spectral weight are further quantified by fitting the photoinduced Coulomb-screening component with the equilibrium spectrum $\alpha(E)$ and its derivatives \cite{kulalov2000}
\[
    \Delta \alpha(E) \approx c_{0}\alpha(E)
+ c_{1}\frac{\partial \alpha}{\partial E}
+ c_{2}\frac{\partial^{2}\alpha}{\partial E^{2}} .
\label{eq:screening_fit}
\]
Here, $c_0$ accounts for changes in spectral intensity, $c_{1}$ captures the spectral shift, and $c_{2}$ describes redistributions of spectral weight. The fitted results displayed in  Fig.~S3, show that photoinduced Coulomb screening mainly leads to a shift of exciton resonance and a weaker redistribution in the spectral weight, while the change in oscillator strength is negligible.

%%%%%%%%%%%%%%%%%%%%%%%%%%%%%%%%%%%%%%%%%%%%%%%%%%%%%%%%%%%%%
%%%%%%%%%%%%%  XTA spectra at the Br K-edge on the ps  %%%%%%%%%%
%%%%%%%%%%%%%%%%%%%%%%%%%%%%%%%%%%%%%%%%%%%%%%%%%%%%%%%%%%%%%
\subsection{XTA spectra at the Br K-edge on the picosecond timescale}

%%%%%%%%%%%%%%%%%%%%%%%%%%  XAS  spectra at ps   %%%%%%%%%%%%%%%%%%%%%%%%%
%

\begin{figure}[H]
    \centering
    \includegraphics[width=1\linewidth]{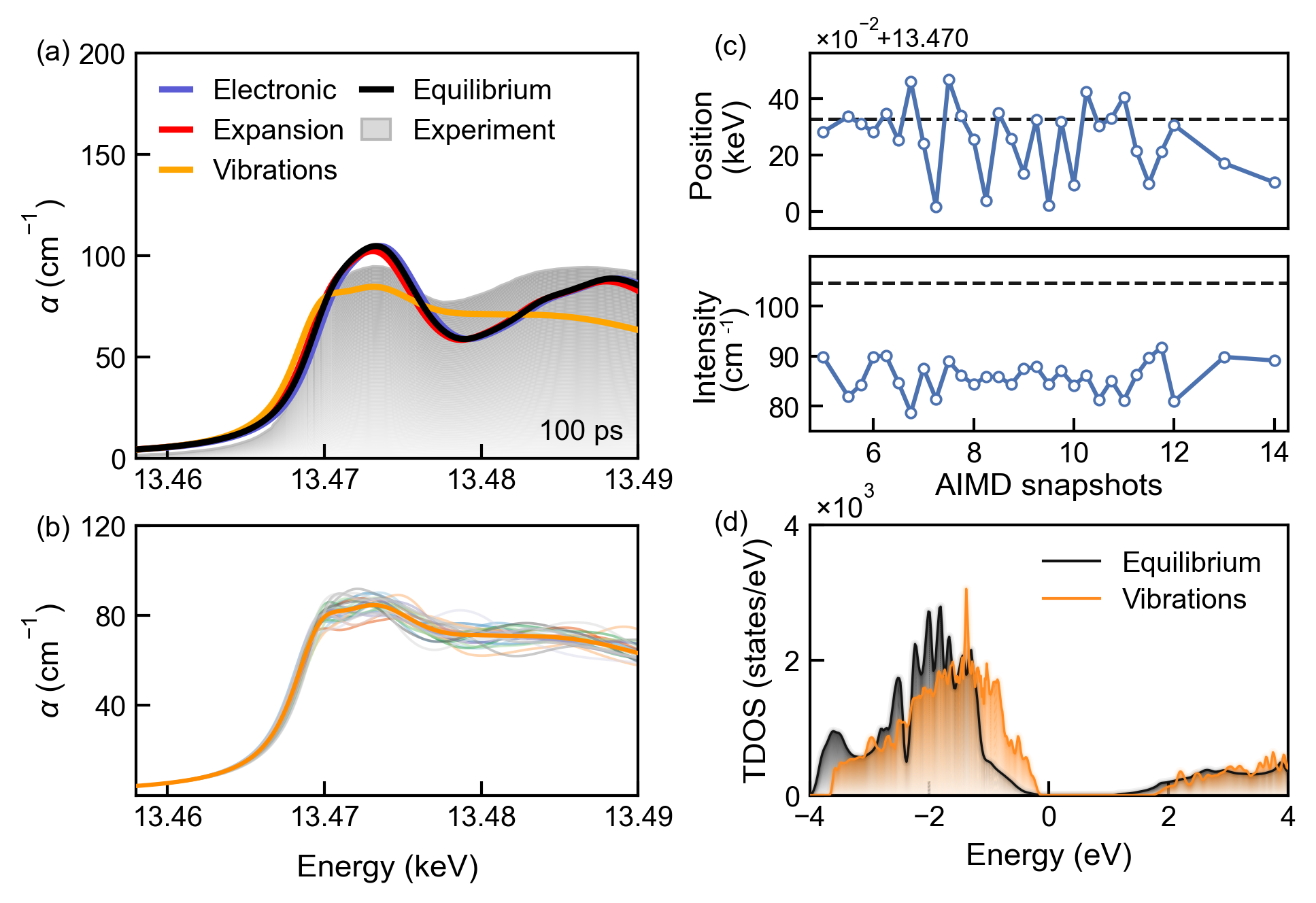}
    \caption{(a) Comparison of the equilibrium (black curve) and the nonequilibrium XAS spectra at the Br K-edge of \ce{CH3NH3PbBr3} at a time delay of \SI{100}{\pico\second} after pump excitation, showing the contributions from photoexcited carriers (blue curve), lattice expansion (red curve), and lattice vibrations (orange curve). The electronic spectrum is calculated at an excitation density of $n_{e}=3.0\times 10^{18}$ cm$^{-3}$. Lattice expansion and vibrations are modeled at 373 K. The experimental spectrum is taken from Ref.~\cite{liu2020}. (b) XAS spectra of AIMD snapshots (colored curves) and their average vibrational spectrum (orange curve). (c) Peak position (top panel) and intensity (bottom panel) extracted from the XAS spectra of representative AIMD snapshots, with the equilibrium values indicated by dashed black lines. (d) Total density of states for the equilibrium structure (black) and a representative vibrational snapshot (orange).}
    \label{fig:vib}
\end{figure}

On the hundreds of picoseconds timescale, the transient responses arise simultaneously from electronic and thermally driven lattice contributions. Figure~\ref{fig:vib}a displays the nonequilibrium XAS spectra at the Br K-edge at a time delay of \SI{100}{\pico\second}, showing the individual contributions from photoexcited carriers (blue curve), lattice expansion (red curve), and vibrations (orange curve), respectively. The electronic spectrum is calculated at an experimental excitation density of $n_e = 3.0 \times 10^{18}$ cm$^{-3}$ \cite{droseros2024}. Lattice expansion and vibrations are modeled at 373 K~\cite{wang2024}. Similar to the femtosecond spectra, the electronic contributions from photoexcited carriers lead to a spectral blueshift of \SI{35}{\milli\electronvolt}. Lattice expansion causes a slight reduction in spectral intensity and a redshift of \SI{46}{\milli\electronvolt}, which can be attributed to the decreased energy gap between the Br 1s core state and the conduction-band states in the expanded structure. To assess the contribution from lattice vibrations, we average the XAS spectra of individual configurations sampled from the AIMD trajectory (colored curves in Fig.~\ref{fig:vib}b) to obtain the vibrational spectrum (orange curve), which shows a redistribution of spectral weight relative to the equilibrium spectrum. 

This redistribution is further quantified by the spectral moments of the vibrational and the equilibrium spectra, as summarized in Table~\ref{tab:spectral_moments}. The zeroth spectral moment, $M_{0}=\int I(E)\, \mathrm{d}E$, evaluates the integrated spectral weight and characterizes the absorption intensity; the first spectral moment, $\bar{E}=\frac{\int E\,I(E)\,\mathrm{d}E}{\int I(E)\,\mathrm{d}E}$, reflects the average excitation energy; the second spectral moment, $\sigma^{2}=\frac{\int (E-\bar{E})^{2} I(E)\,\mathrm{d}E}{\int I(E)\,\mathrm{d}E}$, measures the spread of the spectral weight around the average excitation energy $\bar{E}$, with $\sigma=\sqrt{\sigma^{2}}$ characterizing the spectral width; the third spectral moment, $\gamma_{1}= \frac{\int (E-\bar{E})^{3} I(E)\,\mathrm{d}E} {\left(\int I(E)\,\mathrm{d}E\right)\sigma^{3}}$, describes the spectral asymmetry, where $\gamma_{1}<0$ ($\gamma_{1}>0$) indicates that the spectrum contains low- (high-) energy tails relative to the average excitation energy~\cite{gordon1963}. Details about the spectral moments are provided in Section~S4 of the Supporting Information. Changes in the spectral moments (Table~\ref{tab:spectral_moments}) indicate that the redistribution of spectral weight is dominated by a reduction in absorption intensity and a suppression of spectral asymmetry, whereas change in the average excitation energy is negligible. These findings are further analyzed in Fig.~\ref{fig:vib}c. Compared to the equilibrium value indicated by the dashed black line, the corresponding peak energies in most spectra of the AIMD snapshots fluctuate around the equilibrium value (top panel), while the intensities are generally reduced compared with the equilibrium value (bottom panel). The redistribution of spectral weight due to lattice vibrations can be attributed to a renormalization of the electronic structure. Local structural fluctuations break the lattice symmetry, leading to the mixing and broadening of electronic states and redistributing the density of states over different energy regions, as shown in Fig.~\ref{fig:vib}d.

\begin{table}[H]
\centering
\caption{Spectral moments of equilibrium and vibrational spectra in Fig.~\ref{fig:vib}a, integrated over the \SIrange{13.46}{13.49}{\kilo\electronvolt} range. The spectral moments \(M_{0}\), \(\bar{E}\), \(\sigma\), and \(\gamma_{1}\) quantify the spectral intensity, average excitation energy, spectral width, and spectral asymmetry, respectively. The last column shows the relative change of the spectral moments induced by lattice vibrations with respect to the equilibrium value.
}

\begin{tabular}{lccc}
\hline
Moment & Equilibrium & Vibrational & Relative change (\%) \\
\hline
$M_{0}$ & $1.83\times10^{3}$ & $1.73\times10^{3}$ & $-5.44$ \\
$\bar{E}$ (\si{\electronvolt}) & $13478.3$ & $13477.7$ & $-0.004$ \\
$\sigma$ (\si{\electronvolt}) & $7.06$ & $6.96$ & $-1.40$ \\
$\gamma_{1}$ & $-8.28\times10^{-2}$ & $-6.75\times10^{-2}$ & $-18.4$ \\
\hline
\end{tabular}
\label{tab:spectral_moments}
\begin{flushleft}
\end{flushleft}
\end{table}

%%%%%%%%%%%%%%%%%%%%%%%%%%  XTA  spectra at ps   %%%%%%%%%%%%%%%%%%%%%%%%%

\begin{figure}[H]
    \centering
    \includegraphics[width=0.55\linewidth]{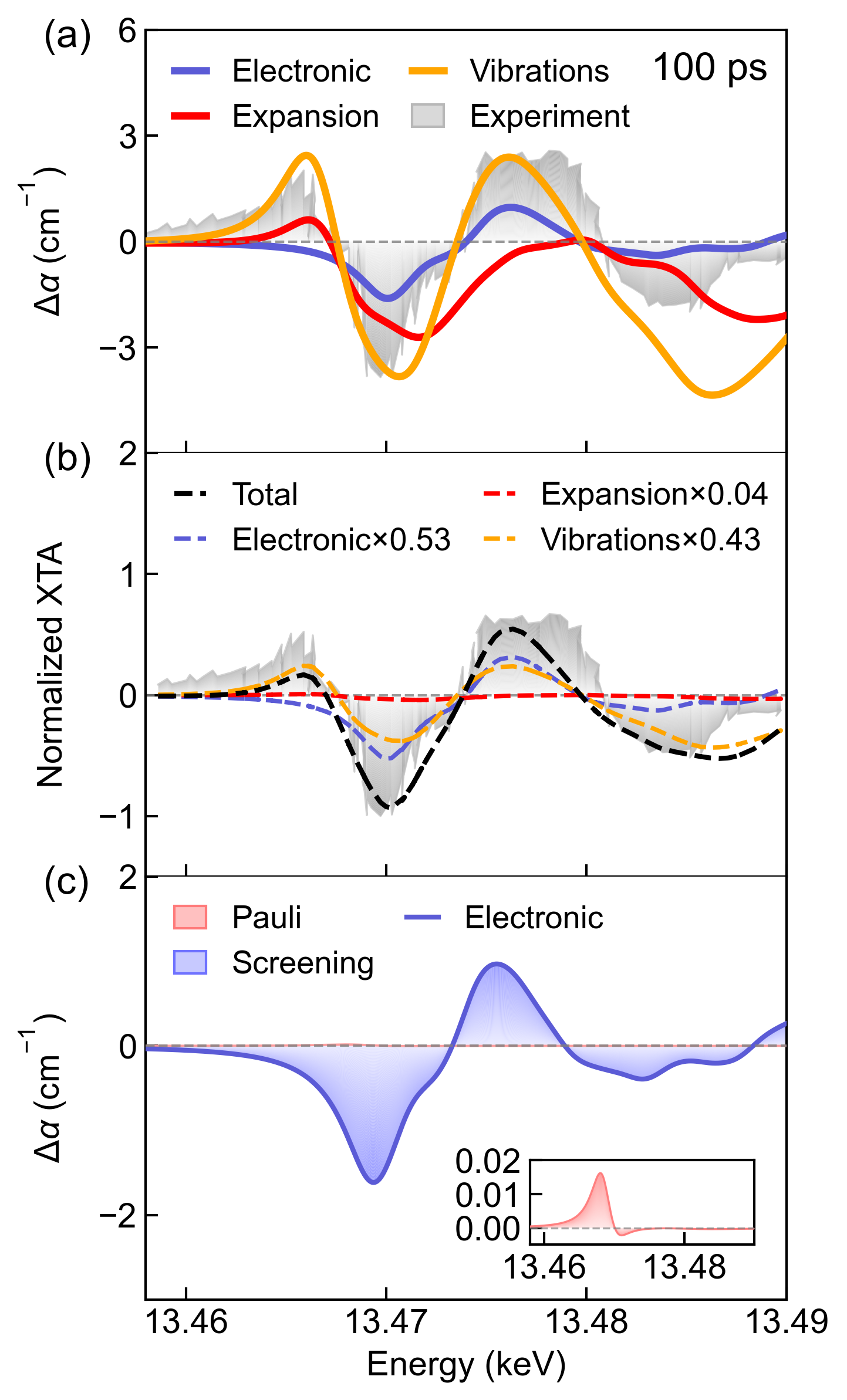}
    \caption{(a) XTA spectra at the Br K-edge of \ce{CH3NH3PbBr3} at a time delay of \SI{100}{\pico\second}, showing individual contributions from photoexcited carriers (blue curve), lattice expansion (red curve), and lattice vibrations (orange curve). The vibrational component is scaled by a factor of 0.2. The experimental spectrum is taken from Ref.~\cite{liu2020}. (b) Fit of the experimental XTA spectrum using the normalized calculated components. The dashed colored curves show the calculated contributions weighted by the fit coefficients of 0.53, 0.04, and 0.43, respectively, and the black dashed curve shows their sum. (c) Decomposition of the electronic XTA spectrum into contributions from photoinduced Coulomb screening (blue area) and Pauli blocking (red area). Inset: Zoom into the Pauli-blocking contribution.
    }
    \label{fig:XTA_ps}
\end{figure}

\newpage

We now disentangle the electronic and lattice contributions to the transient absorption spectra, as shown in Fig.~\ref{fig:XTA_ps}a. The electronic XTA spectrum (blue curve) captures the main derivative-like signature of the experimental counterpart~\cite{liu2020}, in particular the negative feature near $\sim$\SI{13.470}{\kilo\electronvolt} and the positive feature around $\sim$\SI{13.478}{\kilo\electronvolt}. In contrast, the positive pre-edge signal near $\sim$\SI{13.466}{\kilo\electronvolt} is absent. When lattice expansion (red curve) is included, a weak positive pre-edge feature near $\sim$\SI{13.466}{\kilo\electronvolt} emerges. Nevertheless, lattice expansion alone remains insufficient to match the experimental lineshape. Only including lattice vibrations (orange curve) yields satisfactory agreement with experiment, \ie reproducing not only the pre-edge signal but also the overall spectral profile. 

In order to gain a quantitative understanding of the experimental XTA lineshape, we fit it as a linear combination of the calculated spectra, as shown in Fig.~\ref{fig:XTA_ps}b. The experimental XTA lineshape is primarily captured by the electronic and vibrational components, with coefficients of 0.53 and 0.43, respectively, while the lattice-expansion component is negligible. Decomposing the electronic XTA spectrum into photoinduced Coulomb screening (blue area) and Pauli blocking (red area), as shown in Fig.~\ref{fig:XTA_ps}c, we find that the screening contribution gives rise to the derivative-like lineshape, with a negative feature near $\sim$\SI{13.470}{\kilo\electronvolt} and a positive feature near $\sim$\SI{13.478}{\kilo\electronvolt}. We further analyze this contribution using Eq.~\ref{eq:screening_fit} and yield the curves shown in Fig.~S4. The fit shows that a screening-induced blueshift is the dominant electronic effect, while spectral redistribution and intensity reduction are negligible. The Pauli blocking contribution is nearly two orders of magnitude smaller and mainly appears at $\sim$\SI{13.468}{\kilo\electronvolt}. This analysis clarifies that the electronic contribution to the transient response on the picosecond timescale is also dominated by photoinduced Coulomb screening, similar to the femtosecond timescale.

%%%%%%%%%%%%%%%%%%%%%%%%%%%%%%%%%%%%%%%%%%%%%%%%%%%%%%%%%%%%%
%%%%%%%%%%%%%   TA spectra at the Br K-edge on the fs  %%%%%%%%%%
%%%%%%%%%%%%%%%%%%%%%%%%%%%%%%%%%%%%%%%%%%%%%%%%%%%%%%%%%%%%%
\subsection{Optical TA spectra on the femtosecond timescale}

%%%%%%%%%%%%%%%%%%%%%%%%   AS and TA spectra at fs   %%%%%%%%%%%%%%%%%%%%%%%

\begin{figure}[H]
    \centering
    \includegraphics[width=0.5\linewidth]{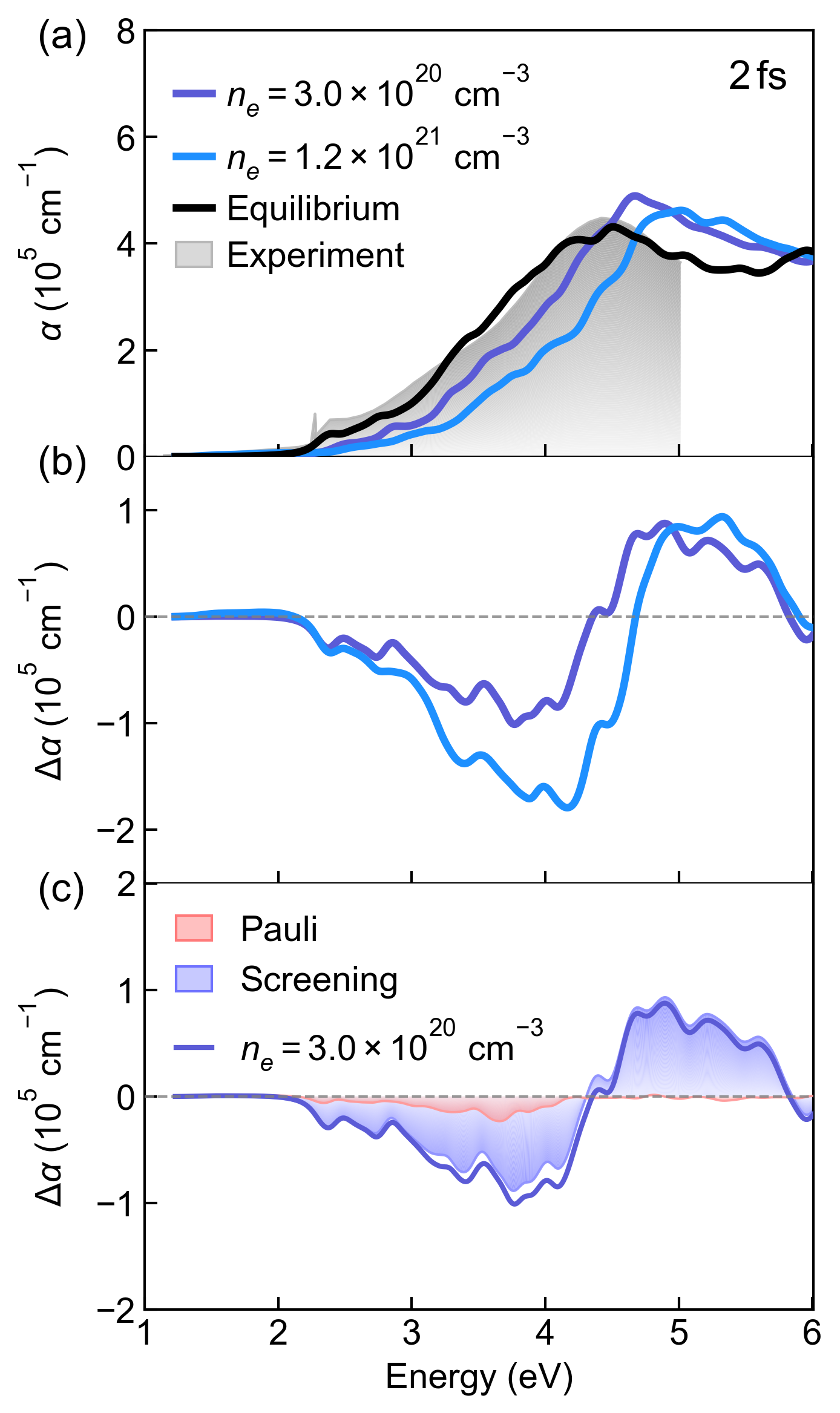}
    \caption{(a) Comparison of the equilibrium (black curve) and nonequilibrium optical absorption of \ce{CH3NH3PbBr3} at different excitation densities $n_e$ (blue curves) and a time delay of \SI{2}{\femto\second}, showing the electronic contributions. The experimental spectrum (gray area) is taken from Ref.~\cite{mannino2020}. (b) Corresponding optical TA spectra at the same time delay. (c) Decomposition of the optical TA spectrum at $n_{e}=3.0\times10^{20}$ cm$^{-3}$ into contributions from photoinduced Coulomb screening (blue area) and Pauli blocking (red area).
}
   \label{fig:optical_fs}
\end{figure}

We now extend our analysis to the optical regime, where the absorption spectrum determines the light-harvesting capability of the material. Figure~\ref{fig:optical_fs} displays that the calculated equilibrium spectrum (black curve) agrees very well with the experimental counterpart (gray area). At a time delay of \SI{2}{\femto\second} after photoexcitation, the optical absorption spectrum (blue curve) exhibits a redistribution of spectral weight, as well as a blueshift of \SI{0.53}{\electronvolt} at $n_e = 3.0 \times 10^{20}$ cm$^{-3}$ and \SI{0.96}{\electronvolt} at $n_e = 1.2 \times 10^{21}$ cm$^{-3}$, respectively. This blueshift gives rise to a derivative-like optical TA lineshape in Fig.~\ref{fig:optical_fs}b, with a negative peak at $\sim$\SI{4.0}{\electronvolt} and a positive peak at $\sim$\SI{4.9}{\electronvolt}. The absence of the isosbestic point at $\sim$\SI{4.5}{\electronvolt} indicates a non-linear behavior of the TA amplitude with excitation density. The TA amplitude at $\sim$\SI{4.0}{\electronvolt} increases approximately as $\sqrt{n_{\mathrm e}}$, similar to the XTA spectra on the femtosecond timescale.

%%%%%%%%%%%%%%%%%%   Decomposition of TA spectra at fs   %%%%%%%%%%%%%%%%%%
Figure~\ref{fig:optical_fs} decomposes the optical TA spectrum into contributions from Pauli blocking and photoinduced Coulomb screening, demonstrating the latter as the dominating factor. As shown by its further decomposition in Fig.~S5, the screening-induced effect shows up predominantly in terms of an energy shift. By contrast, Pauli blocking gives rise to a negative signal in the range of \SIrange{2}{4}{\electronvolt}. This contribution is more substantial than that in the XTA case (Fig.~\ref{fig:XTA_fs}c), indicating that photoexcitation suppresses more optical transitions than core-level transitions, resulting in a more pronounced phase-space filling in the optical region \cite{droseros2024}.

\newpage

%%%%%%%%%%%%%%%%%%%%%%%%%%%%%%%%%%%%%%%%%%%%%%%%%%%%%%%%%%%%%
%%%%%%%%%%%%%  TA spectra at the Br K-edge on the ps   %%%%%%%%%%
%%%%%%%%%%%%%%%%%%%%%%%%%%%%%%%%%%%%%%%%%%%%%%%%%%%%%%%%%%%%%
\subsection{Optical TA spectra on the picosecond timescale}

\begin{figure}[H]
    \centering\includegraphics[width=1\linewidth]{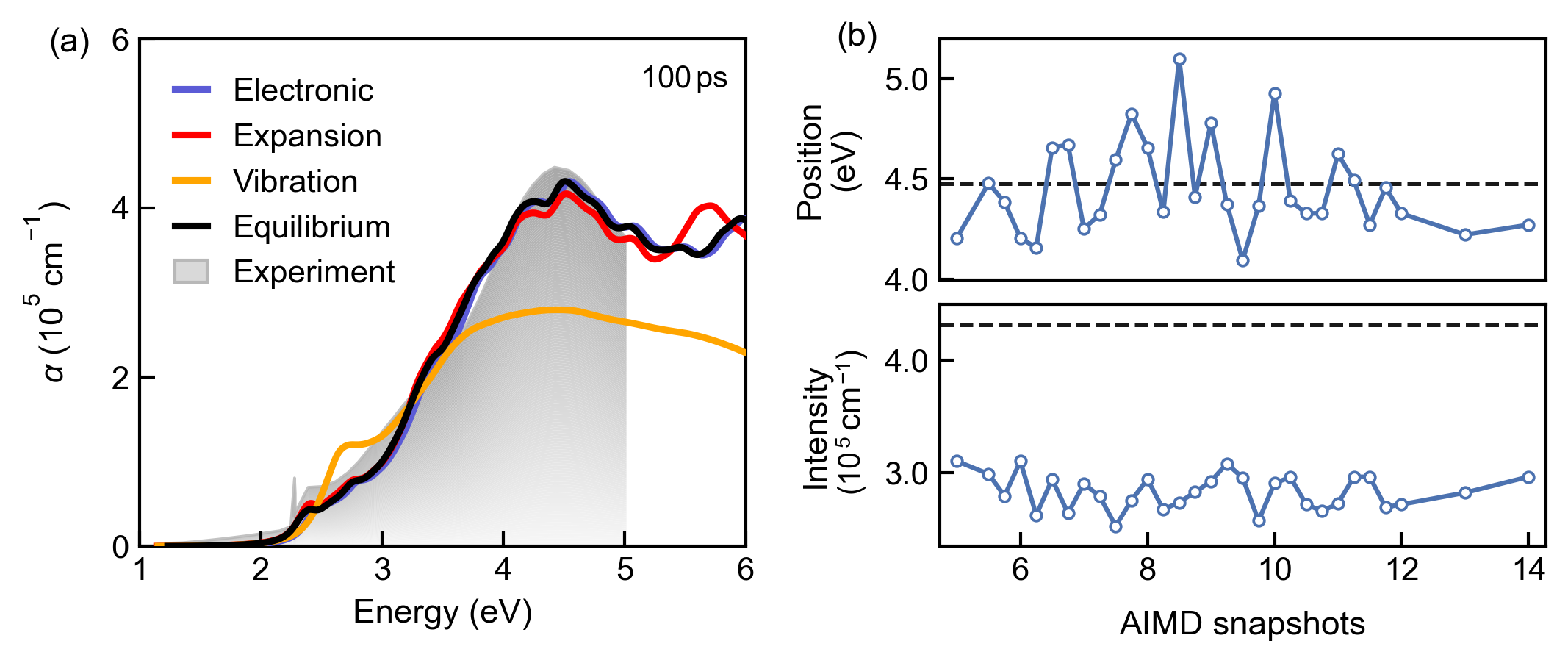}
    \caption{(a) Comparison of equilibrium (black curve) and nonequilibrium optical absorption spectra of \ce{CH3NH3PbBr3} at a time delay of \SI{100}{\pico\second} after pump excitation, showing the contributions from photoexcited carriers (blue curve), lattice expansion (red curve), and lattice vibrations (orange curve). The electronic spectrum is calculated at an excitation density of $n_{e}=3.0\times 10^{18}$ cm$^{-3}$. Lattice expansion and vibrations are modeled at 373 K. The experimental spectrum is taken from Ref.~\cite{mannino2020}. (b) Peak positions (top panel) and intensities (bottom panel) extracted from the optical absorption spectra of representative AIMD snapshots, with the equilibrium values indicated by dashed black lines.}
    \label{fig:optical_AS_ps}
\end{figure}

%%%%%%%%%%%%%%%%%%%%%%%%%%  AS  spectra at ps   %%%%%%%%%%%%%%%%%%%%%%%%%
We now turn to the optical absorption spectra on a longer timescale, \ie a time delay of \SI{100}{\pico\second}, as shown in Fig.~\ref{fig:optical_AS_ps}a. The blueshift of \SI{0.043}{\electronvolt} caused by photoexcited carriers is more than one order of magnitude smaller than that on the femtosecond timescale. Lattice expansion induces a blueshift due to an increase in the band gap (see Fig.~S1). This trend is different from typical inorganic semiconductors \eg \ch{Si}, \ch{GaAs}, and \ch{GaN}, where the band gap narrows with increasing lattice constant \cite{varshni1967,vurgaftman2003}. To examine the effects of lattice vibrations, we calculate the spectral moments of the vibrational spectra, as shown in Table~\ref{tab:optical_spectral_moments}. Compared with the equilibrium case, we observe a redistribution of spectral weight, dominated by a reduction in absorption intensity and a decrease in spectral asymmetry. In addition, the spectrum exhibits a minor redshift and a slight broadening. Figure~\ref{fig:optical_AS_ps}b shows that the AIMD-snapshot spectra exhibit peak positions that fluctuate around the equilibrium value, together with reduced peak intensities relative to the equilibrium spectrum.  

\begin{table}[h]
\centering
\caption{Spectral moments of the equilibrium spectrum and the vibrational spectrum in Fig.~\ref{fig:optical_AS_ps}a, integrated over the \SIrange{1.0}{6.0}{\electronvolt} range. The spectral moments \(M_{0}\), \(\bar{E}\), \(\sigma\), and \(\gamma_{1}\) quantify the spectral intensity, average excitation energy, spectral width, and spectral asymmetry, respectively. The last column shows the relative change of the spectral moments induced by lattice vibrations with respect to the equilibrium value.}
\begin{tabular}{lccc}
\hline
Moment & Equilibrium & Vibrational & Relative change (\%) \\
\hline
$M_{0}$ & $1.05\times10^{6}$ & $8.03\times10^{5}$ & $-23.4$ \\
$\bar{E}$ (\si{\electronvolt}) & $4.56$ & $4.41$ & $-3.20$ \\
$\sigma$ (\si{\electronvolt}) & $0.874$ & $0.935$ & $+6.98$ \\
$\gamma_{1}$ & $-0.316$ & $-0.205$ & $-34.3$ \\
\hline
\end{tabular}
\label{tab:optical_spectral_moments}
\end{table}

\begin{figure}[H]
    \centering\includegraphics[width=0.55\linewidth]{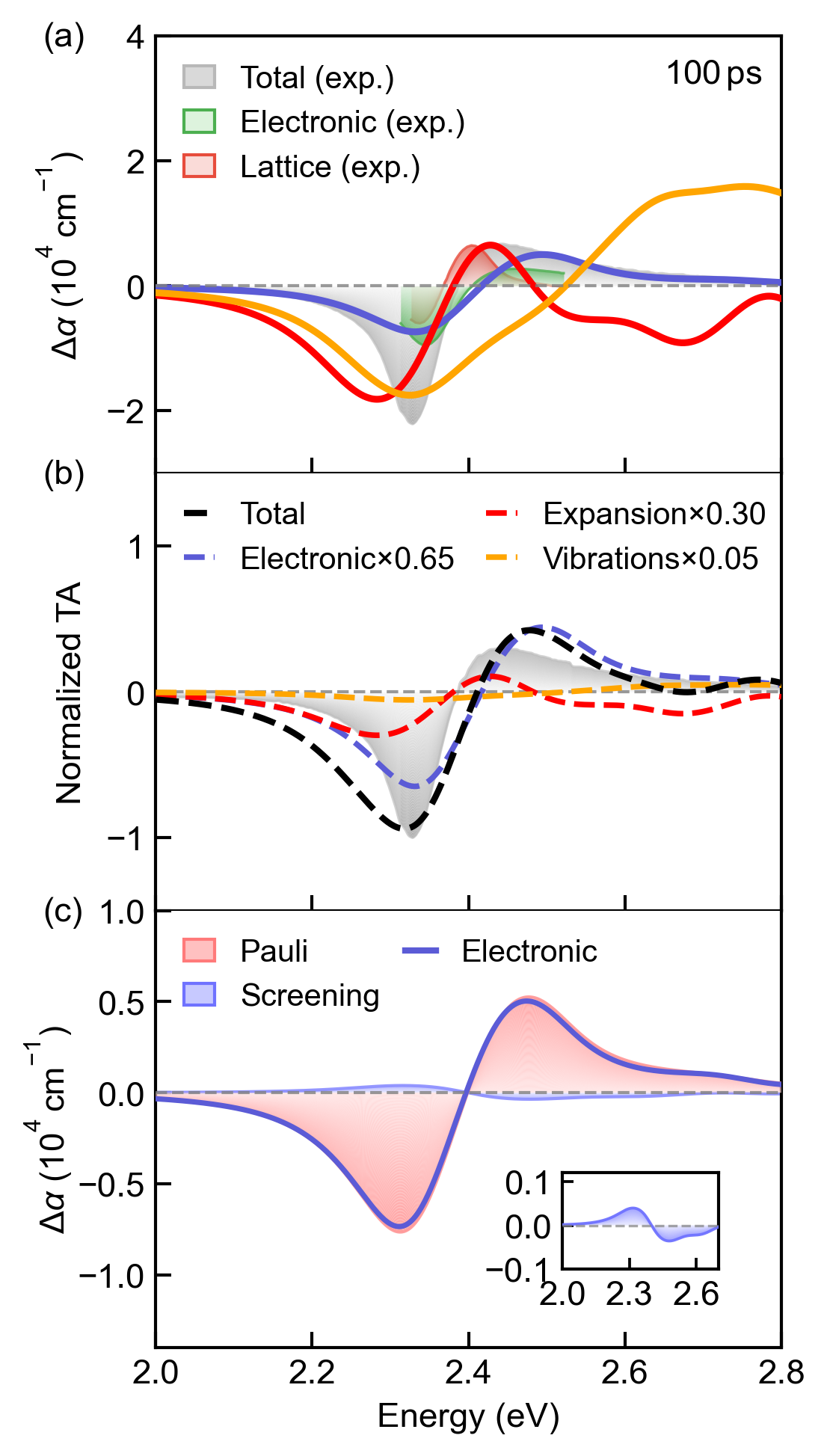}
    \caption{(a) Optical TA spectra of \ce{CH3NH3PbBr3} at a time delay of \SI{100}{\pico\second}, showing the contributions from photoexcited carriers (blue curve), lattice expansion (red curve), and lattice vibrations (orange curve). The experimental spectra (gray, red and green areas) are taken from Ref.~\cite{wang2024}. (b) Fit of the experimental TA spectrum using the normalized calculated components. The dashed colored curves show the calculated components weighted by the fit coefficients of 0.65, 0.30, and 0.05, respectively, and the black dashed curve is their sum. (c) Decomposition of the optical TA spectrum into contributions from photoinduced Coulomb screening (blue area) and Pauli blocking (red area). Inset: Zoom into the contribution from photoinduced Coulomb screening.}
    \label{fig:optical_TA_ps}
\end{figure}

%%%%%%%%%%%%%%%%%%%%%%%%%%  TA  spectra at ps   %%%%%%%%%%%%%%%%%%%%%%%%%
Figure~\ref{fig:optical_TA_ps}a shows the transient spectra in the optical region. The electronic contribution (blue curve) exhibits a symmetric derivative-like lineshape, with a negative feature near $\sim$\SI{2.35}{\electronvolt} and a positive feature around $\sim$\SI{2.48}{\electronvolt}. It is overall close to the experimental electronic spectrum (green area) \cite{wang2024}, but underestimates the negative measured signal around $\sim$\SI{2.35}{\electronvolt}. This inconsistency can likely be attributed to limitations of the Fermi-Dirac distribution in describing the actual carrier distribution generated by the pump at a time delay of \SI{100}{\pico\second}. Lattice expansion (red curve) leads to a negative signal around $\sim$\SI{2.30}{\electronvolt} and a positive signal near $\sim$\SI{2.42}{\electronvolt}, the latter closely matching the experimental thermal spectrum (red area)~\cite{wang2024}. In contrast, the contribution from lattice vibrations (orange curve) exhibits a qualitatively different lineshape. This suggests that the lattice contributions may arise primarily from expansion rather than vibrations. To quantitatively resolve the contributing components, we fit the experimental spectrum (gray area), which includes both electronic and lattice contributions, as a linear combination of the calculated spectra shown in Fig.~\ref{fig:optical_TA_ps}b. This reveals that the spectrum is dominated by the electronic component together with lattice expansion, with coefficients of 0.65 and 0.30, respectively, while the vibrational contribution turns out to be minor. These results show that the optical TA spectrum is sensitive to lattice expansion, in contrast to the XTA spectrum, which is more sensitive to structural fluctuations.

%%%%%%%%%%%%%%%%%%   Decomposition of TA spectra at ps   %%%%%%%%%%%%%%%%%%

Figure~\ref{fig:optical_TA_ps}c disentangles contributions from photoinduced Coulomb screening and Pauli blocking. It shows that the latter dominates the transient response, driving the blueshift of the absorption edge, which is known as the Burstein-Moss effect~\cite{niedzwiedzki2021}. At a time delay of \SI{100}{\pico\second}, the conduction states occupied by excited electrons prohibit further transitions and block absorption at the corresponding energies. In contrast, photoinduced Coulomb screening is negligible. This weak screening response may be attributed to the low excitation densities on the picosecond timescale as well as the weak Coulomb interaction of optical excitons in \ch{CH3NH3PbBr3} \cite{jiang2019}. This behavior differs from the XTA case as well as the femtosecond optical TA spectrum, where photoinduced Coulomb screening is much more pronounced. Overall, these results identify Pauli blocking as the dominant electronic contribution in the optical region at low excitation densities, in agreement with experimental observations that phase-space filling governs the optical transient response in the low-density regime~\cite{droseros2024}.

%%%%%%%%%%%%%%%%%%%%%%%%%%%%%%%%%%%%%%%%%%%%%%%%%%%%%%%%%%%%%%%%%%%%%%%%%%%%%%%%%%%%%
%%%%%%%%%%%%%%%%%%%%%%%%%%%%%%%%%%%%%%%%%%%%%%%%%%%%%%%%%%%%%%%%%%%%%%%%%%%%%%%%%%%%%
\section{Conclusions}
%%%%%%%%%%%%%%%%%%%%%%%%%%%%%%%%%%%%%%%%%%%%%%%%%%%%%%%%%%%%%%%%%%%%%%%%%%%%%%%%%%%%%
%%%%%%%%%%%%%%%%%%%%%%%%%%%%%%%%%%%%%%%%%%%%%%%%%%%%%%%%%%%%%%%%%%%%%%%%%%%%%%%%%%%%%

By combining a nonequilibrium extension of the BSE with AIMD, we establish a general framework to disentangle the electronic and lattice contributions to the pump-probe spectra of photoexcited materials. This approach treats electron and lattice dynamics on equal footing, making it particularly suited for the soft lead halide perovskites studied here. Applying this approach to the X-ray and optical transient absorption spectra of \ce{CH3NH3PbBr3}, we identify the microscopic origins of the transient spectral features and reveal their pronounced dependence on both delay time and probe-energy window. At a time delay of \SI{2}{\femto\second}, both X-ray and optical TA spectra arise from electronic contributions, while lattice contributions are negligible. Specifically, photoinduced Coulomb screening governs both spectral regions by weakening the effective electron-hole attraction, thus reducing exciton binding energies and producing a nonlinear blueshift of the excitonic resonances with increasing excitation density. Pauli blocking, in contrast, is negligible in the XTA spectrum but becomes appreciable in the optical region. At a delay of \SI{100}{\pico\second}, lattice contributions become essential and compete with the remaining electronic contributions. The XTA spectra are governed primarily by lattice vibrations and photoinduced Coulomb screening, whereas the optical TA spectra are dominated by lattice expansion and Pauli blocking. 

These results demonstrate that transient absorption spectra cannot generally be interpreted as direct fingerprints of carrier dynamics alone. Instead, their lineshapes reflect the time- and energy-dependent interplay among carrier occupations, Coulomb screening, lattice expansion, and vibrations. Our work therefore provides a quantitative route for assigning the microscopic origins of transient spectral features and establishes a connection between ultrafast spectroscopic observables and the underlying electronic and structural responses of photoexcited materials.

\section{Associated Content}

\subsection{Data Availability Statement}
All input and output files are available in the NOMAD data infrastructure with the following link:\\https://doi.org/10.17172/nomad.m1qz-2fcr

\section{Author Information}
\subsection{Notes}
The authors declare no competing financial interest.

\section{Acknowledgments}
Work supported by the German Research Foundation through the priority program SPP2196 Perovskite Semiconductors, project No. 424709454. L.Q. acknowledges financial support from the Berlin Equal Opportunities Programme for Women and thanks Benedikt Maurer for valuable discussions. The authors gratefully acknowledge the computing time made available to them on the high-performance computer "Lise" at the NHR center NHR@ZIB. This center is jointly supported by the Federal Ministry of Research, Technology, and Space and the state governments participating in the NHR (www.nhr-verein.de). 

%%%%%%%%%%%%%%%%%%%%%%%%%%%%%%%%%%%%%%%%%%%%%%%%%%%%%%%%%%%%%%%%%%%%%
%% If you are using classical BibTeX rather than biblatex,
%% remove the \printbibliography and uncomment the \bibliograpy one
%%%%%%%%%%%%%%%%%%%%%%%%%%%%%%%%%%%%%%%%%%%%%%%%%%%%%%%%%%%%%%%%%%%%%

%\printbibliography
%\bibliography{acs-main,references}
\bibliography{references}

\end{document}